\documentclass{aa}  

\usepackage{graphicx}
\usepackage{hyperref}
\usepackage{txfonts}
\usepackage{lipsum}
\usepackage{subcaption}         
\usepackage{lscape}             
\usepackage{placeins}           
\usepackage{color,soul}

\makeatletter
\@ifundefined{makeLineNumber}{}{%
  \renewcommand\makeLineNumber{\relax}%
  \providecommand\thelinenumber{}%
}
\makeatother

\begin{document}
\nolinenumbers

   \title{Definition and Realization \\of the International Lunar Reference Frame}



   \author{K. Sośnica\inst{1}
        \and A. Fienga\inst{2}
        \and D. Pavlov\inst{3}
        \and N. Rambaux\inst{4}
        \and R. Zajdel\inst{1,5}
               }

   \institute{Institute of Geodesy and Geoinformatics, Wroclaw University of Environmental and Life Sciences, Norwida 25, 50-375 Wrocław, Poland 
             \email{krzysztof.sosnica@upwr.edu.pl} \\
            \and Observatoire de la Côte d'Azur, Université Côte d'Azur, CNRS, IRD, 250 avenue A. Einstein 06560 Valbonne, France\\
            \and
            St. Petersburg Electrotechnical University, Faculty of Computer Science and Technology, ul. Professora Popova 5, 197022 St. Petersburg, Russia \\
            \and Sorbonne Universit\'e, Observatoire de Paris, Universit\'e PSL, CNRS, Laboratoire Temps Espace, LTE, F-75014 Paris, France \\
            \and
            Research Institute of Geodesy, Topography and Cartography, Geodetic Observatory Pecný, Ústecká 98, Zdiby, 25066, Czechia
            }

   \date{Received October 10, 2025}

 
  \abstract
   {All future lunar missions require a definition of the lunar reference system and a realization in the form of the lunar reference frame to ensure consistent products for positioning, navigation, cartography, and timing. }
   {This paper defines the origin, orientation, and scale of the Lunar Reference System (LRS), as well as provides numerical solutions for the first realization of the International Lunar Reference Frame (ILRF). ILRF is defined as the Principal Axis (PA) system, attached to the surface and co-rotating with the Moon, with its origin in the lunar center of mass (lunocenter).} 
   {The ILRF realization is based on variance component estimation of the three lunar ephemeris solutions: INPOP21a, DE430, and EPM2021 for the series of the position of the lunar center of mass and rotation Euler angles – precession, nutation, and proper rotation. The solution is valid starting with the period covered by Lunar Laser Ranging (LLR) data in 1970 and ending with extrapolated ILRF realizations in 2052 for future lunar missions. }
   {The combined ILRF is characterized by the mean error of 17.6 cm for 2010-2030, where 15.3 cm comes from the origin and 8.6 cm from the orientation realization. The error in the realization of the origin is mainly caused by a poor geometry of the retroreflector network, resulting in a high correlation between the scale and the X component of the lunocenter in PA. The LLR post-fit residuals in ILRF are at the level of 2-3 cm in terms of the standard deviations of one-way ranges for best-performing LLR stations. The mean errors of the transformation between ILRF and other reference frame realizations in PA are at the level of 3 cm, whereas the mean transformation error to the DE421 Mean Earth frame equals 5 cm. }


   \maketitle

\section{Introduction}

The rapid growth of lunar exploration initiatives --- from NASA's Artemis program, ESA’s Moonlight and NovaMoon, to numerous commercial and national missions --- highlights the need for a common reference infrastructure for the Moon, including reference frames and time systems \citep{Ventura2025, Iess2025}. Precise positioning, navigation, cartography, and timing services require a standardized lunar reference system and its practical realization in the form of a reference frame. Without such a common framework, inconsistencies between data products, navigation solutions, and archived cartographic information could hinder interoperability between missions and agencies.

\subsection{Lunar reference frame realizations: PA vs. ME}

The development of lunar reference systems and frames builds upon several decades of research in celestial mechanics, ephemerides, and geodetic observations. Early analytic theories of the Moon’s motion, such as the ELP2000-82 ephemeris \citep{Chapront1983}, provided a foundation for selenographic cartography but were gradually replaced by high-precision numerical ephemerides. The availability of Lunar Laser Ranging (LLR) since 1969 revolutionized lunar geodesy, allowing the determination of the Moon’s orbit, rotation, and tidal dissipation with centimeter precision \citep{Dickey1994}.

Two primary body-fixed realizations of the lunar reference system have historically been used. The Principal Axis (PA) frame, defined by the diagonalization of the Moon’s inertia tensor, co-rotates with the Moon and is well-suited for navigation and positioning because its axes are dynamically defined and directly tied to physical librations \citep{Folkner2014, Rambaux2025}. Consequently, PA frames are direct by-products of lunar ephemerides. In contrast, the Mean Earth (ME) frame is based on the mean direction of the Earth and the mean rotation axis of the Moon and has been widely adopted for cartography, mapping, and archiving of selenographic data, including Lunar Reconnaissance Orbiter (LRO/LOLA) and historical lunar maps. The ME frame also ensures consistency with long-standing cartographic conventions. As the definition of the ME frame is deduced from lunar ephemerides and their corresponding PA frames, recent discussions emphasize the adoption of the PA frame for operational navigation and positioning while retaining the ME frame for legacy cartography and archiving. Transformation procedures between PA and ME frames are now well established \citep{Rambaux2025}, but a unified, internationally recognized realization of a lunar reference frame is still missing.

\subsection{Lunar ephemerides}

Modern realizations of lunar reference frames are based on high-precision numerical ephemerides that assimilate decades of LLR observations and spacecraft tracking data. The three main series used internationally include:

\begin{itemize}
    \item The Development Ephemeris (DE) series (JPL, USA), e.g., DE430: A high-precision planetary and lunar ephemeris \citep{Standish1982} constructed with extensive use of LLR and spacecraft tracking, modeling tidal dissipation in the Earth--Moon system, a fluid lunar core with core--mantle coupling, and relativistic formulations \citep{Folkner2014}, which are used for many NASA missions \citep{Park2021}.

\item The INPOP series (IMCCE-Nice Observatory, France): fitted to LLR, spacecraft tracking (radio and Very Long Baseline Interferometry, VLBI) and optical observations, incorporating detailed planetary ephemerides, relativistic corrections, and Earth and lunar tides \citep{Fienga2020}. INPOP21a has been employed for independent tests of General Relativity (GR) and alternative theories of gravity \citep{Fienga2024a}. INPOP10a and INPOP19a are used by ESA for GAIA and JUICE mission operations, while INPOP will be applied for data analysis and archiving of the BepiColombo mission.

\item The Ephemerides of Planets and the Moon (EPM) series (IAA RAS, Russia): the latest realization, EPM2021 integrates long LLR series and planetary observations, having the same lunar model as DE430, but with independently fit parameters of gravitational field, lunar interior, and kinematic corrections \citep{Pitjeva2014, Pavlov2016}.

\end{itemize}
Although each of these ephemerides achieves centimeter-level accuracy in fitting LLR radial ranges, systematic discrepancies of several decimeters remain in the position components of the lunar center-of-mass, and up to several centimeters in orientation (Euler angles). These discrepancies reflect differences in background dynamical models, such as tidal dissipation laws, lunar interior structure, treatment of the lunar core, parameterization of kinematic corrections, observational data selection and weighting, and initial conditions. As a result, while each ephemeris is internally consistent, its realizations of the lunar reference frame are not strictly interoperable. Comparative studies \citep{Williams2013,  Viswanathan2019} have documented systematic differences between these ephemerides, particularly in the lunar center-of-mass position and orientation, underscoring the need for a combined solution. A review of the three ephemerides for planets can be found, e.g., in \cite{Fienga2024a}, whereas comparisons between the lunar ephemerides with accuracy assessments are given in \cite{Fienga2024b}.

\subsection{Motivation and the goal of this study}
The International Astronomical Union (IAU) has recently adopted Resolution II (2024), establishing the Lunar Celestial Reference System (LCRS) and proposing the Lunar Coordinate Time (TCL), mirroring the terrestrial framework \citep{IAU2024, Petit2010, Altamimi2023}. However, the IAU resolutions define only the conceptual system, not attached to the lunar surface; whereas a practical, combined realization of a lunar reference frame has not yet been delivered. Currently, agencies and researchers rely separately on DE, INPOP, or EPM solutions, which leads to inconsistencies of up to tens of centimeters or meters -- unacceptable for future high-precision lunar navigation, geodesy, and science.

The International Association of Geodesy, together with the IAU, established in 2023 a Joint Working Group (JWG) 1.1.3 on Lunar Reference Frames to address the issues of missing reference standards for future lunar missions. One of the main tasks of the JWG is to provide proposals and practical solutions for the lunar reference frames, time, and height systems. 

The niche addressed by this paper is therefore the absence of a combined, internationally recognized realization of a lunar reference frame that averages over and reconciles the differences between the major ephemerides. By combining DE430, INPOP21a, and EPM2021 using variance component estimation (VCE), this study delivers the first realization of the International Lunar Reference Frame (ILRF). The ILRF provides a statistically robust, decimeter-level realization of the Lunar Reference System (LRS), aligned with the PA frame, with quantified origin and orientation errors, and transformation parameters to existing ephemerides and ME-based frames. This combined ILRF is designed to serve as a common reference standard for navigation, cartography, and timing in the new era of lunar exploration.

\section{Definitions}

Following the IAU resolutions from 2024 \citep{IAU2024} that define the quasi-inertial LCRS, we propose a reference system and a reference frame co-rotating with the Moon. The definitions are consistent, to the possible extent, with the definitions of the International Terrestrial Reference System (ITRS) from the Conventions of the International Earth Rotation and Reference Systems Service, IERS \citep{Petit2010} and its realizations in the form of a series of International Terrestrial Reference Frames, ITRFs \citep{Altamimi2023}.

The Lunar Reference System (LRS) is defined in such a way that:
\begin{itemize}
    \item its origin coincides with the center of mass of the Moon (lunocenter),
\item its orientation is defined by the diagonal matrix of the lunar tensor of inertia; hence, the system co-rotates with the Moon,
\item its scale follows the GR framework as defined by the IAU 2024 resolutions.
\end{itemize}

The realization of the LRS, the International Lunar Reference Frame (ILRF), is obtained in such a way that:
\begin{itemize}
    \item its origin is based on the combination of INPOP21a, DE430, and EPM2021, using VCE and referred to the mean center of mass of the Earth,    
    \item its orientation is provided by three Euler angles from the VCE combined solution based on INPOP21a, DE430, and EPM2021. 
    \item its scale is realized complementarily with the origin and orientation in a consistent GR framework,
    \item the weights for the origin and orientation derived from VCE are the same for all components.
\end{itemize}

LLR is the technique used for the origin, orientation, and scale realization in each contributing ephemeris solution; however, the background models, geophysical parameters, and fundamental constants may differ in each contributing constituent. Nevertheless, each contribution includes its own best fit of the associated parameters to the LLR observations. The series of lunocenter positions is given in the Geocentric Celestial Reference System (GCRS), where the XY-plane is the mean equator of J2000.0, the X-axis lies in that plane and points toward the vernal equinox of J2000.0, the Z-axis points toward the celestial north pole, and the Y-axis completes the right-handed system. The ILRF realization is consistent with the PA system in terms of orientation. However, transformations to other ephemeris expressed in PA and ME (and vice versa) are possible using the transformation parameters that are provided along with the ILRF (see Section 6: Transformation parameters). 

The combined ILRF covers the period between 1970 and 2052, which includes the historical high-quality LLR observations and extrapolation in time to support future lunar missions. However, the quality and reliability of the extrapolated period are not as good as for the periods with real LLR observations. Accordingly, an update of the ILRF realization is foreseen as soon as new lunar retroreflectors or longer time series of LLR observations or new observation techniques are available. For instance, new observation techniques and continuous measurements from the Earth that have been proposed by ESA in the framework of the NovaMoon lander \citep{Ventura2025} may contribute to the improvement of future ILRF realizations.

ILRF is distributed as a time series of positions, velocities, orientation angles, and their first derivatives (rates) in ASCII format sampled every 0.75 days, as well as in the form of Chebyshev polynomials in an analogy to the distribution of other ephemerides. The positions and orientation parameters are provided in Barycentric Dynamical Time (TDB). The linear rates, along with ephemeris, can be used for the transformation to other time systems, such as the TCL, which has recently been proposed for the Moon and future lunar missions and whose definition and realization are still the subject of discussion. Hence, TDB is employed in the current ILRF realization, which is consistent with the time scale used for planetary ephemerides, whereas future releases will be provided in TCL as defined by the Bureau International des Poids et Mesures (BIPM). The orientation parameters in ILRF are expressed as three Euler angles, which are not corrected by the empirically derived kinematic parameters. However, the kinematic parameters are recommended for high-accuracy applications; thus, they are provided and discussed in Section 4.3. 

\section{Combination strategy}

The combination strategy follows a modified version of VCE  \citep{Pukelsheim1976}, which is widely used for deriving combined solutions in satellite geodesy, deriving combined orbits in the International GNSS Service \citep{Zajdel2025}, and for combining different geodetic parameters \citep{Schaffrin1981}.  \citet{Beutler1995} proposed the orbit combination for the first time for artificial Earth satellites and proved that the combined orbit satisfies the equation of motion provided that the weights in the combination are constant and the differences of combined orbits are small. Hence, we apply the combination strategy in the analogous way to \citet{Beutler1995} with the constant weight values derived from VCE.

We assume that three datasets (EPM2021, INPOP21a, DE430) provide lunar reference frame parameters at discrete epochs $t\in T_{1970\div2052}$ with a 18 h-sampling (0.75 days), resulting in $n$-epochs. Each dataset $ i \in ({1,2,3}) $ provides a 6D state vector:
   \begin{equation}
    \label{eq1}
\mathbf{X}_i\left(t\right)=\left[\begin{matrix}\begin{matrix}x_i\left(t\right)\\y_i\left(t\right)\\\end{matrix}\\z_i\left(t\right)\\\begin{matrix}\varphi_i\left(t\right)\\\theta_i\left(t\right)\\\psi_i\left(t\right)\\\end{matrix}\\\end{matrix}\right]\ ,
   \end{equation}
where $x, y, z$ are the lunar origin coordinates, and $\varphi,\theta, \psi $ are orientation parameters. Assuming equal $a priori$ variance (unit covariance), the initial mean estimate is:
   \begin{equation}
    \label{eq2}
\bar{\mathbf{X}}\left(t\right)=\frac{1}{3}\sum_{i=1}^{3}{w_i\mathbf{X}_\mathbf{i}\left(t\right),}
   \end{equation}
where $w_i = 1$ are unit weights in the first step of iteration. 
We estimate residuals for each contributing ephemerides series:
\begin{equation}
\mathbf{r}_i = \mathbf{X}_i - \bar{\mathbf{X}}.
\end{equation}

Then, we split the residual vector into position and orientation (rotation) parts:
\begin{equation}
\mathbf{r}_i(t)\ =\ \left[\begin{matrix}\mathbf{r}_{i, \ pos}\ (t)\\\mathbf{r}_{i,\ rot}\ (t)\\\end{matrix}\right]\ \ \end{equation} 
and we define scalar residual norms to unify units for the position and orientation parts:
\begin{equation}
\begin{matrix}
{d_i\left(t\right)=||\mathbf{r}}_{i, \ pos}\left(t\right)||, \ \ \ \ \ \ \ \\
{\omega_i\left(t\right)=C\ \cdot||\mathbf{r}}_{\mathbf{i},\ rot}\left(t\right)||,\ 
\end{matrix}
\end{equation}
where $C$ is the conversion factor, i.e., the lunar equatorial radius (1,738,000 m) multiplied by the radian conversion factor $\pi/180^{\circ}$. As a result, both the origin and the orientation parameters are expressed in meters. Then, we estimate the total variance $\sigma_i^2$ for dataset i for the origin and normalized orientation:
\begin{equation}
\sigma_i^2 = Var(d_i) + Var(\omega_i),
\end{equation}
where
\begin{equation}
Var(d_i)=\frac{1}{n-1}\sum_{j=1}^{n}{d_{i,j}}^2\ \textrm{and} \ Var(\omega_i)=\frac{1}{n-1}\sum_{j=1}^{n}{\omega_{i,j}}^2,\ \ 
\end{equation}
which allows us to compute weights $w_i$  and normalized weights ${\widetilde{w}}_i$:
\begin{equation}
w_i=\frac{1}{\sqrt{\sigma_i^2}},\ \ {\widetilde{w}}_i=\frac{w_i}{\sum_{k=1}^{3}w_k}.
\end{equation}
Using the normalized weights ${\widetilde{w}}_i$, the combined solution per epoch is:
\begin{equation}
\mathbf{X}_{combined}\left(t\right)=\sum_{i=1}^{3}{{\widetilde{w}}_i\mathbf{X}_\mathbf{i}\left(t\right)}.
\end{equation}

The procedure of weight adjustment is repeated iteratively until convergence, defined as a maximum normalized weight difference of $10^{-15}$, thus, smaller than 1 mm on the lunar surface. The $\mathbf{X}_{combined}\left(t\right)$ from eq. 9 replaces $\bar{\mathbf{X}}\left(t\right)\ $from eq. 3 and the whole procedure is repeated. The same normalized weights ${\widetilde{w}}_i$ are finally applied to the rates of origin (velocities) and the rates of orientation parameters, which also consist of the 6D state vector. However, the rates do not contribute to the weight estimation. 

\section{Combination results}

\begin{table}[h!]
\centering
\caption{Normalized weights $\widetilde{w}_i$ derived from VCE for different lunar ephemeris models.}
\label{tab1}
\begin{tabular}{lc}
\hline\hline 
Ephemeris & VCE weights \\
\hline 
EPM2021   & $0.451454072137127$ \\
INPOP21a  & $0.380949596178373$ \\
DE430     & $0.167596331684500$ \\
\hline\hline 
\end{tabular}
\end{table}

The normalized weights provided by VCE with the combined contribution from the orientation and origin are given in Table \ref{tab1}. The large number of digits in Table \ref{tab1} is dictated by the need to keep sub-mm precision of the combination of the lunar origin in GCRS. The model with constant weight (instead of time-variable) and the same weighting factors for orientation and origin was selected as the simplest, and simultaneously, the most robust solution. \citet{Beutler1995} showed that constant weights are needed to ensure that the combination satisfies the equation of motion for celestial bodies. Other combination options were also tested, e.g., a separation of the weights for the orientation and origin; the adjustment of the scale to one selected model; considering the contribution from the rates of origin and orientation to the weighting factors; or using a different set of contributing ephemerides. However, none of these solutions was characterized by a better performance in terms of LLR residuals; therefore, the simplest and the most robust solution was selected for generating the combined ILRF.

\subsection{Origin}

The origin of the ILRF is expressed with respect to the origin of the ITRF, which is placed in the mean long-term linear center-of-mass of the Earth as sensed by Satellite Laser Ranging (SLR) observations to geodetic satellites \citep{Pearlman2019}. Despite that different ITRF realizations are characterized by slightly different positions of the reference frame origin, all recent ITRF realizations can be considered sufficiently consistent for the ILRF realization. The difference of the ITRF origin does not exceed 3 mm since the ITRF2005, which is also valid for ITRF2008, ITRF2014, and ITRF2020 \citep{Altamimi2023}. We can thus infer that the error emerging from the ITRF origin realization is negligible for the purpose of the ILRF realization. 

\begin{figure*}[h!]
   \centering
   \includegraphics[width=0.85\hsize]{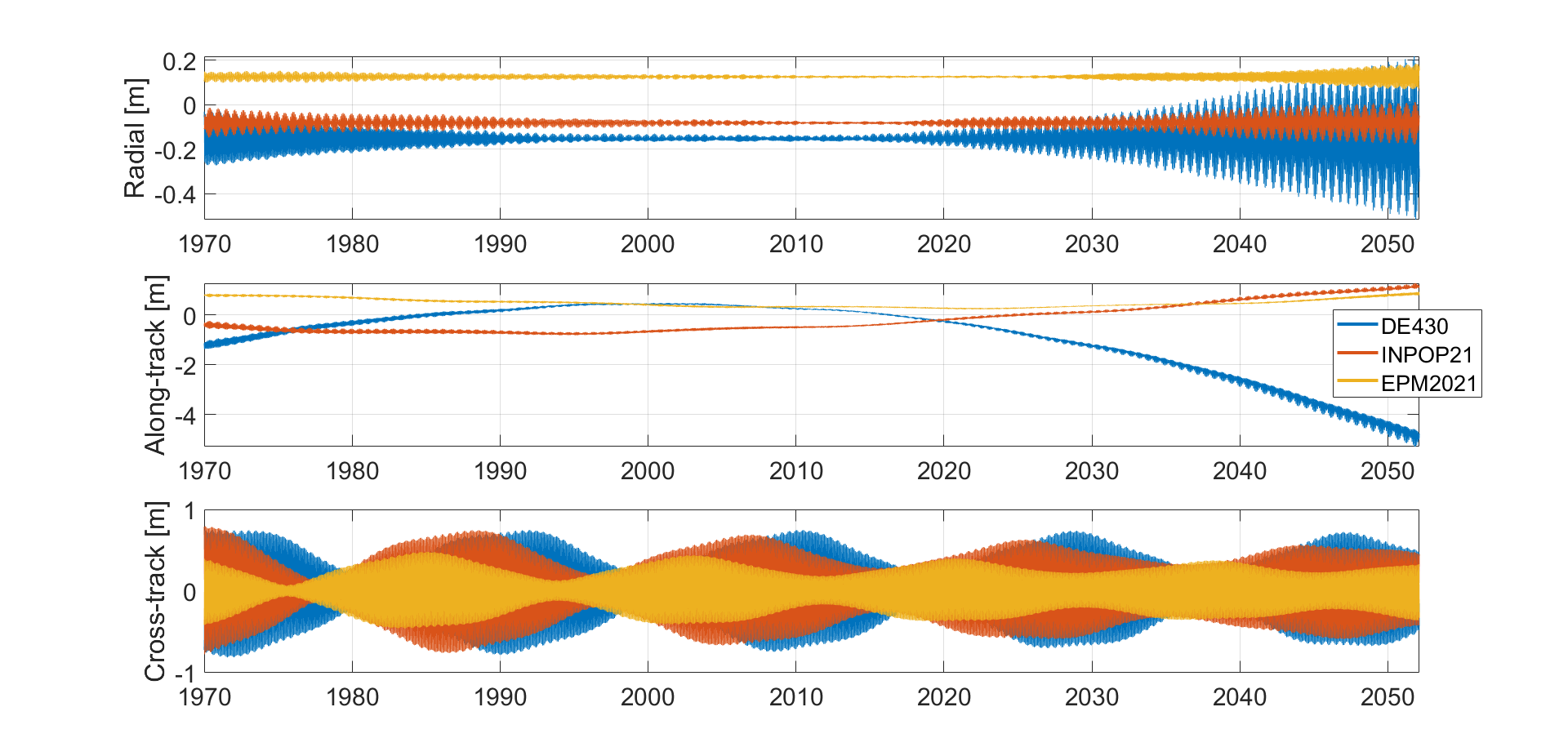}
      \caption{The time series of the ILRF origin realization shown as differences between ILRF and the contributing ephemeris, decomposed into radial, along-track, and cross-track components in GCRS. }
         \label{fig1}
   \end{figure*}

The same SLR stations used for the ITRF origin realization are used for ILRF realization as they also provide LLR observations: Grasse (7845), Matera (7941), Wettzell (8834), and formerly McDonald (7080) and Haleakala (7120). The only station that provides LLR observations but is not included in ITRF realizations due to the lack of SLR observations to LAGEOS satellites is Apache Point Observatory Lunar Laser ranging Operation (APOLLO, 7045). Nevertheless, the realization of ILRF keeps the high level of consistency with the ITRF realizations via common techniques and methods used for deriving the datum-defining parameters. However, during the fit of the lunar ephemerides, the positions and the velocities of the SLR/LLR stations are also fitted and are slightly different from those in ITRF. Each data processing center providing lunar ephemeris estimates derives a different set of station coordinates and velocities, which may lead to some inconsistencies for the integrated terrestrial-lunar reference frames. Nevertheless, the differences in LLR station coordinates are assumed to be an order of magnitude smaller than the differences between lunar ephemerides (Fig. \ref{fig1}). 

The Cartesian coordinate system of GCRS is used in the ILRF distribution; however, the decomposition into radial, along-track, and cross-track components is better suited for the discussion of orbit differences and perturbations than the Cartesian system for all natural and artificial satellites. In GCRS, the differences along the X and Y axes are difficult to interpret because their orientation changes rapidly. The decomposition into radial ($R$), along-track ($S$), and cross-track ($W$) for the $i$-th model reads as:
\begin{equation}
R_i(t)=\frac{{\bar{\mathbf{X}}}_{pos}\left(t\right)\cdot{\ \mathbf{r}}_{i,pos}\left(t\right)}{||{\bar{\mathbf{X}}}_{pos}\left(t\right)||},
\end{equation}

\begin{equation}
S_i(t)=\frac{{\bar{\dot{\mathbf{X}}}}_{pos}\left(t\right)\cdot{\ \mathbf{r}}_{i,pos}\left(t\right)}{||{\bar{\dot{\mathbf{X}}}}_{pos}\left(t\right)||},\ \ 
\end{equation}

\begin{equation}
W_i(t)=\left(\frac{{\bar{\dot{\mathbf{X}}}}_{pos}\left(t\right)\times{\bar{\mathbf{X}}}_{pos}\left(t\right)}{||{\bar{\dot{\mathbf{X}}}}_{pos}\left(t\right)||\cdot||{\bar{\mathbf{X}}}_{pos}\left(t\right)||}\right)\ \cdot{\ \mathbf{r}}_{i,pos}\left(t\right),\ \ 
\end{equation}
where ${\bar{\mathbf{X}}}_{pos}\left(t\right)$ and ${\bar{\dot{\mathbf{X}}}}_{pos}\left(t\right)$ denote the position and velocity vectors of the lunar center-of-mass in GCRS at $t$ from the ILRF, and ${\mathbf{r}}_{i,pos}\left(t\right) $ includes the differences of the position component at the epoch t for the the $i$-th contributing model: DE430, INPOP21a, and EPM2021 with respect to the combined ILRF.

Figure \ref{fig1} illustrates the differences between ILRF and DE430, INPOP21a, and EPM2021 for the lunar center-of-mass positions with respect to the Earth’s center-of-mass decomposed into the radial, along-track, and cross-track constituents. The offset in the radial component describes concurrently the scale offset between contributing models and the combination. The scale difference may result from a slight difference in the background parameters, e.g., the Earth/Moon mass ratio, the Moon's mass, or a different scale alignment to ITRF. The mean scale offset between ILRF and contributing models equals $-15.3$, $-8.2$, and $12.6$ cm for DE430, INPOP21a, and EPM2021, respectively. All solutions are characterized by very small noise below 1 cm in the period 1990-2020, i.e., with the high-accuracy LLR data. The noise tends to increase outside this time window. However, the differences for the radial component are small, with the RMS at the level of 8 cm without removing the mean offset and 1 cm after removing the mean offset for the best-performing model (see Table \ref{tab2}). 

\begin{table}[h!]
\centering
\caption{RMS and STD (no offset) of lunar ephemeris solutions in radial, along-track, and cross-track directions w.r.t. ILRF.}
\label{tab2}
\subcaption*{RMS [m]}
\begin{tabular}{lccc}
\hline\hline 
Ephemeris & Radial & Along-track & Cross-track \\
\hline
EPM2021   & 0.126 & 0.511 & 0.214 \\
INPOP21a  & 0.084 & 0.599 & 0.350 \\
DE430     & 0.168 & 1.666 & 0.372 \\
\hline\hline 
\end{tabular}

\vspace{1em}

\subcaption*{STD, no offset [m]}
\begin{tabular}{lccc}
\hline\hline 
Ephemeris & Radial & Along-track & Cross-track \\
\hline
EPM2021   & 0.012 & 0.171 & 0.214 \\
INPOP21a  & 0.021 & 0.568 & 0.350 \\
DE430     & 0.069 & 1.422 & 0.372 \\
\hline\hline 
\end{tabular}
\end{table}

The cross-track differences are stable at the same level for the entire time series, which means that the orbital plane is well-defined over long time spans. The mean cross-track offsets are equal to zero for all ephemerides. The RMS of cross-track differences is, however, larger than for the radial component, and ranges between 32 and 37 cm (Table \ref{tab2}).

The along-track component is characterized by the largest errors, especially in the predicted part that is not constrained by real LLR observations. The residuals reach even 5 m in 2052 for the DE430 model. Therefore, future ILRF updates are necessary to keep the high accuracy of the reference frame realization, whereas the extension of the ILRF far beyond 2052 does not guarantee any superior accuracy of its realization. The mean RMS values for EPM2021 and INPOP21a models in the along-track component are at the level of 50 cm for the entire period. As a result, the along-track component is affected by the largest errors, substantially exceeding the total contribution from all other origin and orientation components. Small differences in the background constants or models, such as external torques, may systematically accumulate in the along-track component that almost coincides with the Moon's velocity vector. Large discrepancies for the DE430 model explain the smallest weight assigned by VCE for this model (see Table \ref{tab1}) because the errors in the along-track origin component play a key role in the total error budget when deriving the weights. The DE430 model also includes a shorter series of LLR data, ending in 2012 because it is older than INPOP21a and EPM2021, fitted up to 2020 and 2021, respectively. Please note that LLR provides observations that are mostly sensitive to the radial component, whereas the lateral along-track and cross-track components are characterized by an accuracy of two orders of magnitude poorer than the directly observed radial ranges. Therefore, some other observational techniques sensitive to the lateral components, e.g., the VLBI transmitter proposed on NovaMoon, could possibly improve the ILRF realizations \citep{Klopotek2018}. The radial component is determined with the best accuracy because it is well constrained by principles of celestial mechanics and LLR data.

\begin{figure}[h!]
   \centering
   \includegraphics[width=0.8\hsize]{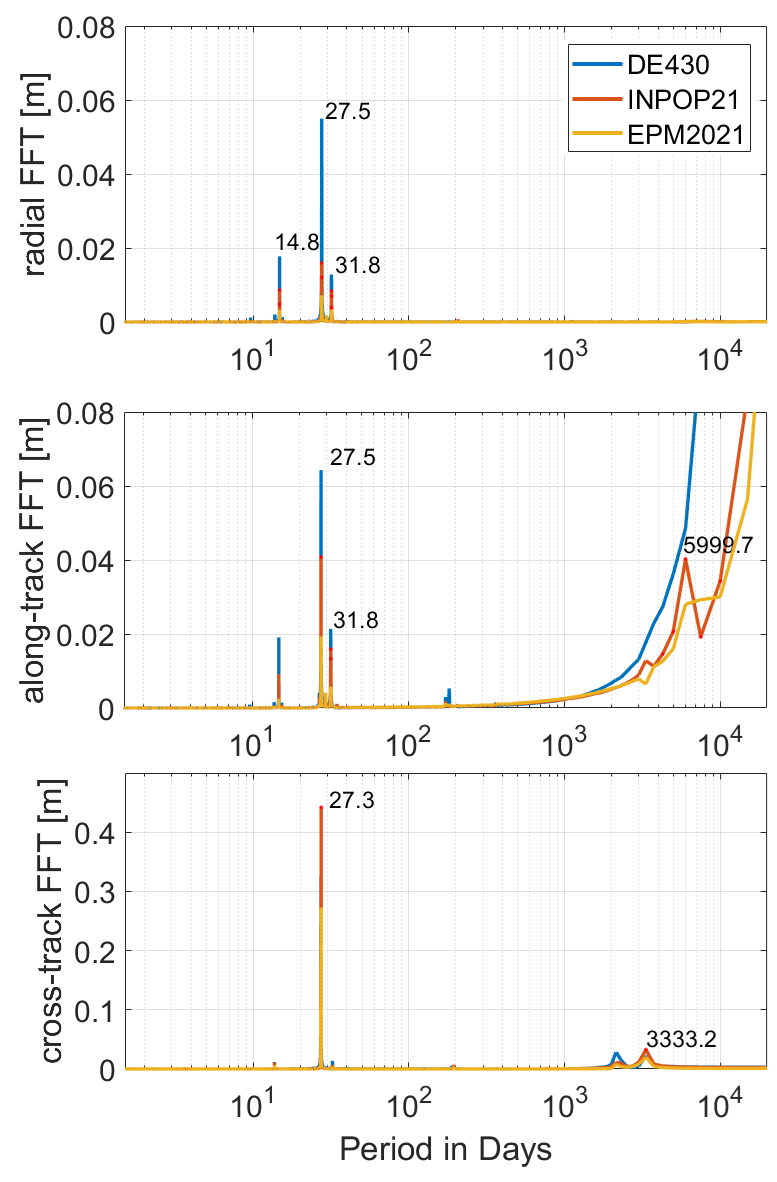}
      \caption{The spectral analysis of the ILRF origin realization differences between ILRF and the contributing ephemeris, decomposed into radial, along-track, and cross-track components.}
         \label{fig2}
   \end{figure}

Figure \ref{fig2} shows the spectral analysis of the differences between ILRF origin components and the contributing ephemeris. For the radial and along-track components, three periods dominate: 14.8, 27.5, and 31.8 days with amplitudes for DE430 of about 2, 5-7, and 2 cm, respectively. For INPOP21a and EPM2021, these differences are smaller and reach up to 4 and 2 cm, respectively. These periods correspond to the semilunar (half-synodic) period, the anomalistic month (the time between successive pericenters), and the evection (the largest monthly perturbation of the Moon’s ecliptic longitude caused by the Sun), for 14.8, 27.5, and 31.8-day periods, respectively. The along-track component also includes the very long perturbations, i.e., in the combination of the anomalistic period of 8.85-year cycle and the draconic period of 18.6 years, resulting in the beat period of 16.9 years (about 6170 days, see Fig. \ref{fig2}).

The largest observed periodic differences are for the cross-track component at the 27.3-day period, with the amplitudes of 44, 33, and 27 cm for INPOP21a, DE430, and EPM2021, respectively. This period corresponds to the sidereal month, i.e., the lunar revolution around Earth relative to the distant stars. The sidereal month is most affected by tides and dissipation, i.e., the differences in core–mantle boundary friction and dissipation mechanisms. Interestingly, for the cross-track component, the largest differences are observed for INPOP21a, and not for DE430, as opposed to the other components. Other identified periods with the amplitudes of 2-3 cm include 5.9 years (which results from a coupling of the 8.85-year apsidal precession and the 18.6-year nodal regression) and about 8.85 years. These differences can be explained by the additional 8 years of observations used by INPOP21a and EPM2021. 

Please note that the same period of 27.3 days as for the cross-track component is identified in the spectral analysis for all Euler components (see Section 4.2), which indicates the large sensitivity of the cross-track to the rotation parameters. A small rotation, i.e., a difference in orientation of the Moon, produces a displacement of any surface point mostly in the cross-track direction. The orientation that defines those rotations is referenced to inertial space; therefore, its fundamental period is the sidereal month of 27.3217 days. The instantaneous Earth–Moon distance corresponding to the radial component varies as the Moon moves around its elliptical orbit; those distance variations repeat once per anomalistic month (27.554 d), whereas the along-track couples to the radial component because the differences are produced by small differences in orbital mean motion and are also sensitive to eccentricity-related terms, which carry strong power at the anomalistic frequency. The radial and along-track differences emerge from different dynamical models, such as core coupling, tidal dissipation laws, lunar gravity coefficients taken from different sources or fit to different datasets, and different LLR station corrections in INPOP21a, DE430, and EPM2021. On the contrary, the cross-track differences correspond to the differences in libration modeling. 

\subsection{Orientation}

   \begin{figure*}[h!]
   \centering
   \includegraphics[width=0.75\hsize]{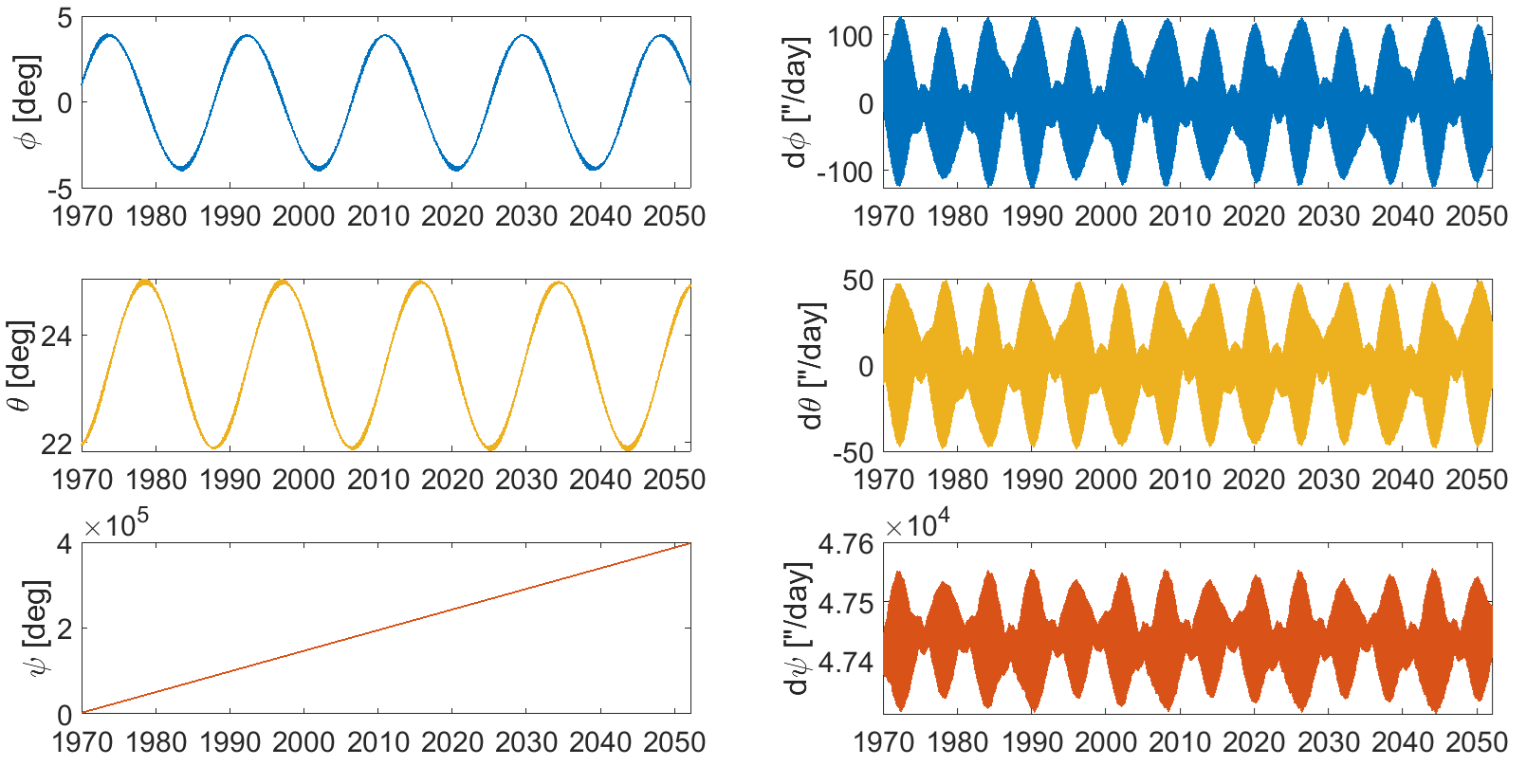}
      \caption{Orientation parameters – Euler angles of the ILRF realization (left) and their rates (right). }
         \label{fig3}
   \end{figure*}

      \begin{figure*}[h!]
   \centering
   \includegraphics[width=0.85\hsize]{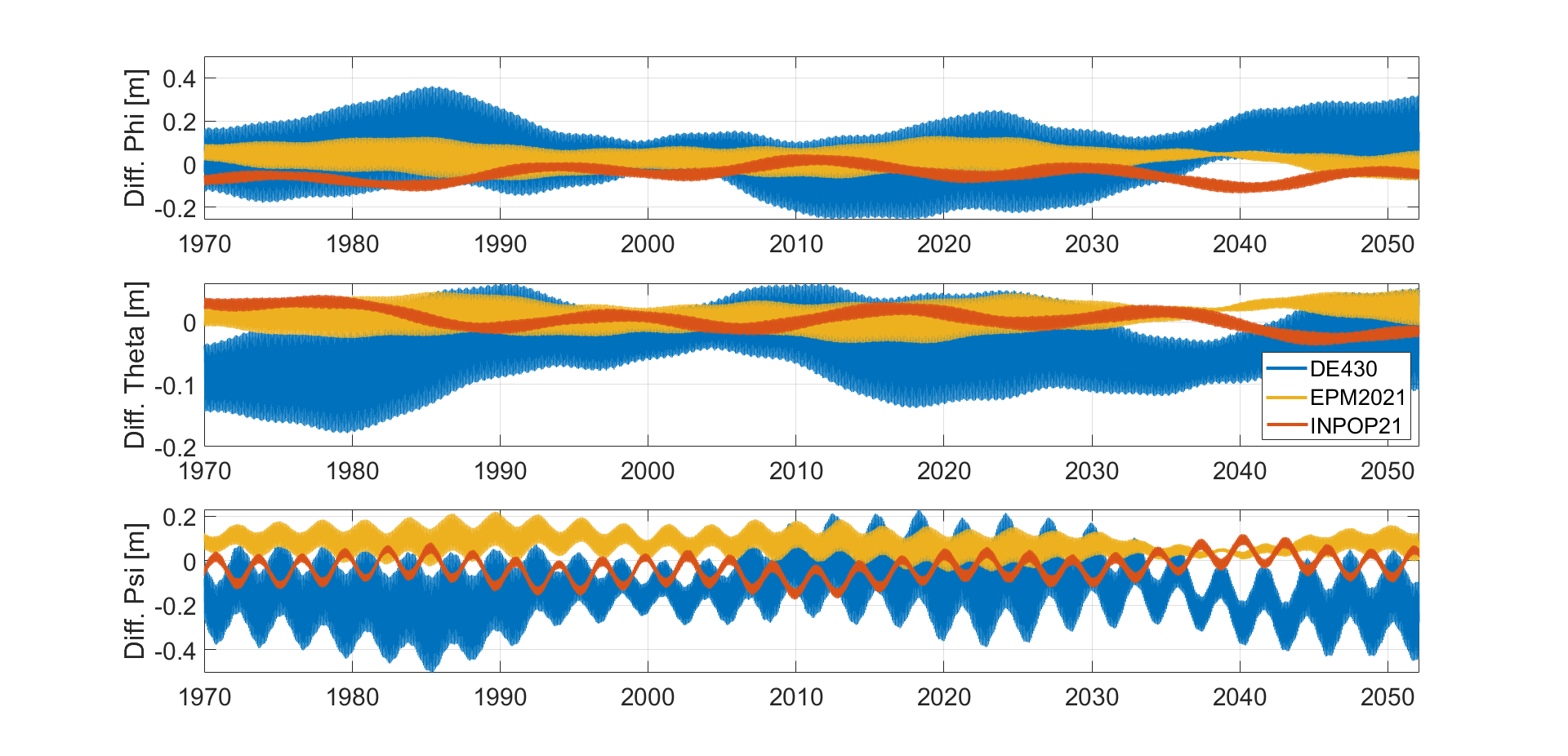}
      \caption{Differences of the orientation parameters between the combined ILRF and individual solutions: DE430, EPM2021, and INPOP21a expressed in meters on the lunar surface. }
         \label{fig4}
   \end{figure*}

The transformation between quasi-inertial LCRS and body-fixed LRS in the PA system is described through the rotation (Euler) angles – precession, nutation, and proper rotation ($\varphi, \theta, \psi$) with respect to the Lunar Celestial Reference Frame (LCRF) in TDB. The orientation of the LCRF is consistent with the International Celestial Reference Frame; however, its origin coincides with the Moon’s center of mass (as described in Section 4.1). The transformation to the lunar-fixed PA from the quasi-inertial LCRF reads as:
   \begin{equation}
PA=\ R_Z\left(\psi\right)R_X\left(\theta\right)R_Z\left(\varphi\right)\ LCRF
   \end{equation}
with $R_i$ corresponding to the rotation matrix around the $i$-th axis.

Figure \ref{fig3} shows the time series of three Euler angles and their rates for the combined ILRF. The precession and nutation parameters follow the libration periodic pattern, whereas the proper rotation is dominated by the secular term. Please note that for the weight calculation, only the orientation parameters and the lunar center-of-mass series were considered, whereas the rates did not contribute to the weights. However, the same weights are applied to derive both the parameters and their rates. The reference epoch for $\psi$ is set up as the first epoch of the ILRF realization, i.e., 01.01.1970. The values of the proper rotation $\psi$ in ILRF are consistent with the EPM2021 and DE430, whereas they are shifted by $204\cdot360^{\circ}$ w.r.t. INPOP21a due to a different selection of the initial epoch for $\psi$ in INPOP21a (J2000 for INPOP, whereas for DE and EPM the reference epoch is the average epoch of the fit period). 

\begin{table}[ht]
\centering
\caption{The RMS values of the differences w.r.t. ILRF for the orientation parameters. 
The smallest values for each component are shown in bold.}
\label{tab3}
\begin{tabular}{lccc}
\hline\hline
RMS [m]   & $\varphi$ & $\theta$ & $\psi$ \\
\hline
EPM2021   & \textbf{0.054} & 0.022 & 0.092 \\
INPOP21a  & 0.059 & \textbf{0.017} & \textbf{0.063} \\
DE430     & 0.139 & 0.067 & 0.189 \\
\hline\hline
\end{tabular}
\end{table}

Figure \ref{fig4} shows the time series of differences between the combined ILRF and individual solutions: DE430, EPM2021, and INPOP21a. The differences are expressed in meters on the lunar surface to keep the same units as for the origin parameters. Table \ref{tab3} provides the RMS values of differences shown in Fig.~\ref{fig4}. Definitely, the smallest differences are obtained for nutation ($\theta$) with the RMS value of 1.7 cm for INPOP21a. This parameter uniquely describes the rotation around the X-axis, whereas the precession ($\varphi$) and proper rotation ($\psi$) describe the rotation around the Z-axis. Thus, $\varphi$ and $\psi$ are correlated by construction because the mean value of $\theta$ is close to $23^{\circ}$; thus, the separation of these parameters is not complete. However, the amplitudes of the nutation variations are also the smallest out of all orientation parameters (cf. Fig. \ref{fig3}). The RMS values of the precession ($\varphi$) and proper rotation ($\psi$) parameters are at the level of 5-6 cm. No clear temporal deterioration in the orientation is visible for the extrapolation period with no LLR data. Instead, the quality of the combined ILRF w.r.t. the best individual models is at the level of 8-9 cm in terms of 3D RMS for the entire time span. We may thus infer that the ILRF and different ephemeris models offer the Euler angles with consistency at the sub-decimeter level, which is considered sufficient for current and future lunar missions. 

   \begin{figure}[h!]
   \centering
   \includegraphics[width=0.75\hsize]{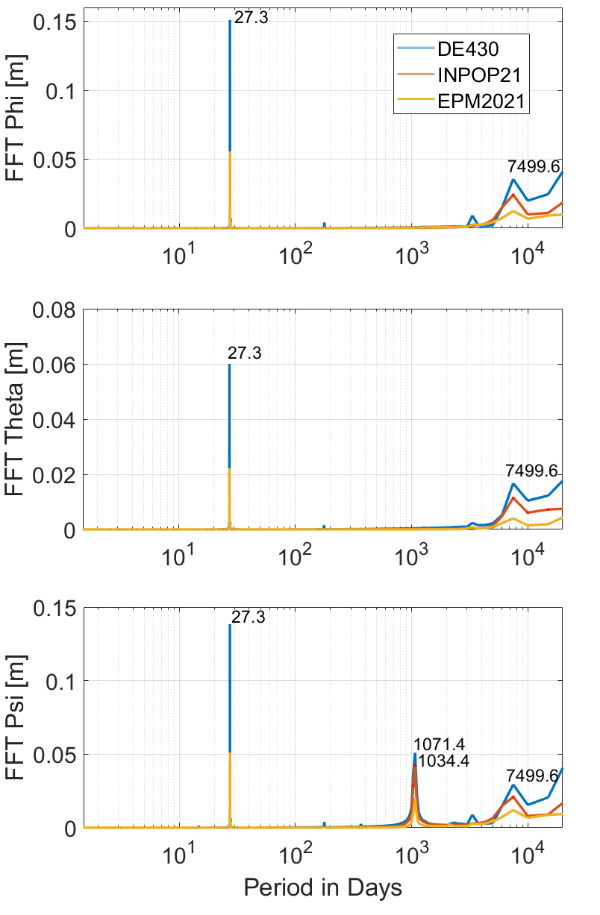}
      \caption{Spectral analysis of differences of the orientation parameters between the combined ILRF and individual solutions: DE430 (blue), EPM2021 (yellow), and INPOP21a (orange) based on Fast Fourier Transform (FFT).}
         \label{fig5}
   \end{figure}

The achieved accuracies of the prediction for Moon orientation are spectacular when compared to the prediction of the Earth orientation. The best methods of Earth orientation predictions for the UT1-UTC parameter provide errors of 0.40 ms after 10 days, which corresponds to 20 cm on the Earth's equator (based on the IERS Second Earth Orientation Parameters Prediction Comparison Campaign from 2022; \cite{Kur2022}). For the Moon, the prediction of the orientation for 30 years can be characterized by a superior accuracy (see Table \ref{tab4}) when compared to the best prediction methods of the Earth orientation just after 10 days (on the surface of a celestial body). This is mostly related to the unpredictable part of the Earth’s atmosphere, oceans, land hydrology, and other geophysical processes that do not appear on the Moon.

Figure \ref{fig5} shows the amplitude spectra of the Euler differences from Fig. \ref{fig4} for each contributing reference frame. The largest differences occur for each component at the period of 27.3 days with differences up to 15, 6, and 14 cm for  $\varphi,\ \theta,$ and $\psi$, respectively for DE430. Differences up to 4 cm can also be identified for the period of about 7500 days, i.e., 20.53 years. This period is related to the libration pattern of the Moon’s rotation relative to Earth (8.85 y and 18.6 y periods) combined with the annual motion of the Sun (1 y), specifically tied to the rotation of the line of apsides. The 20.53 y period corresponds to the repeating orientation of the Moon’s figure, and its orbit relative to the Earth–Sun line that involves integer and half-integer linear combinations of the three fundamental frequencies (1 y, 8.85 y, and 18.6 y). The amplitude of the 20.53 y signal does not, however, exceed 4~cm for all rotation components.

The distinct third signal of the period of about 1071 days can be identified, but this time only for the proper rotation ($\psi$).\  The amplitude of the 1071-day signal is up to 5 cm for DE430. This signal can be explained as the third-harmonic component of the Moon’s physical libration – the apsidal (pericenter) precession period of 8.85 years. The period is also close to 1056 days, which is the free mode of rotation. These three main signals cause most of the observed periodic differences, whereas the amplitudes of all other signals are below 1 cm. The differences with the INPOP and EPM ephemerides give similar behaviors with mainly the same periods in the spectra illustrated in Fig. \ref{fig5}. Additional periods specific to one ephemeris have amplitudes smaller than 1 cm.

\begin{table*}[ht]
\centering
\caption{Mean errors derived as the weighted mean error of the ILRF realization for sub-periods and the entire time series. The 1-$\sigma$ values are provided.}
\label{tab4}
\begin{tabular}{lccccc}
\hline\hline
Mean Error [m] & 1970--1990 & 1990--2010 & 2010--2030 & 2030--2052 & Whole \\
\hline
Origin      & 0.284 & 0.225 & 0.153 & 0.467 & 0.305 \\
Orientation & 0.085 & 0.061 & 0.086 & 0.068 & 0.082 \\
Total       & 0.296 & 0.233 & 0.176 & 0.472 & 0.316 \\
\hline\hline
\end{tabular}
\end{table*}

\subsection{Amplitudes of kinematic corrections and Love number}

To fully benefit from the high-quality ILRF, the so-called kinematic corrections must be applied. These kinematic corrections are due to a lack in the orientation modeling and produce signals for 3 identified periods for which the amplitudes are fitted in order to reduce the differences between the LLR observations and the computed distances (residuals). Without these corrections, the residuals are at the decimeter level, whereas the LLR accuracy is supposed to be centimetric. For the combined ILRF as well as for DE430, INPOP21a and EPM2021, the three main terms are provided: $A_1$, $A_2$, and $A_3$, to maximize the consistency with the contributing models:
   \begin{equation}
\mathrm{\Delta\psi}\ =\ A_1\cos{\left(l^\prime\right)}+\ A_2\cos{\left(2l-2D\right)}+\ A_3\cos{\left(2F-2l\right)},\ \ \ 
   \end{equation}
where $l'$, $2l - 2D$, and $2F - 2l$ are Delaunay arguments which correspond to about 1 year, 206 days, and 3-year periods, respectively. $l'$ denotes the Sun mean anomaly (progress of the Earth-Moon system along its orbit around the Sun), $l$ is the Moon’s mean anomaly (progress of the Moon along its orbit around Earth), $D$ is the mean elongation of the Moon from the Sun (the angular distance between the Moon and the Sun as seen from Earth), and $F$ is the Moon’s argument of latitude (related to the inclination of the lunar orbit relative to the ecliptic). The kinematic corrections emerge from some geophysical processes inside the Moon that are still not fully explained and described, and thus, require empirical corrections. 
For ILRF, the obtained values of kinematic corrections are equal to $A_1 =4.4$, $A_2=1.6$, $A_3=1.2$~mas. 

Different lunar ephemerides include different numbers of kinematic corrections. Increasing the number of empirical kinematic corrections improves the fit of the lunar ephemeris to LLR data; however, the combination of Euler angles can become problematic when not all kinematic corrections are fully specified and provided. The number of kinematic corrections in DE440 has been increased by a factor of two with respect to the previous models, such as DE430. Hence, using DE440 instead of DE430 for the combined ILRF leads to differences up to 2 m for $\phi$ and $\varphi$ and 0.5 m for $\theta$, whereas DE430 is more consistent with EPM2021 and INPOP21a with differences up to 40 cm for all Euler angles (see Fig. \ref{fig2}). Therefore, DE430 has been selected for the combination, instead of DE440. 

Finally, the VCE-combined Love number describing the tidal deformation of the Moon on its surface in the radial direction equals $h_2=0.0432$ for ILRF. 

\subsection{Mean errors of ILRF realization}

Table \ref{tab4} provides the mean errors of the ILRF combination for individual epochs as derived from VCE, with a distinction for the origin and orientation constituents. The smallest combination errors are for the period 2010-2030 with a total error of 17.6~cm, which is dominated by the origin error of 15.3 cm with a contribution from orientation of 8.6 cm. For the whole assessed period, the total error equals 31.6 cm and is dominated by the origin error of 30.5 cm, mostly related to the uncertainties in the along-track components in the prediction period or for the initial period with LLR data of inferior quality. The period 2010-2030 is of special interest for current and future missions; therefore, the total error below 20 cm (1-$\sigma$) should satisfy the needs of most of the lunar-related activities.

\section{Validation}
\subsection{LLR validation without kinematic corrections}

   \begin{figure}[h]
   \centering
   \includegraphics[width=1.00\hsize]{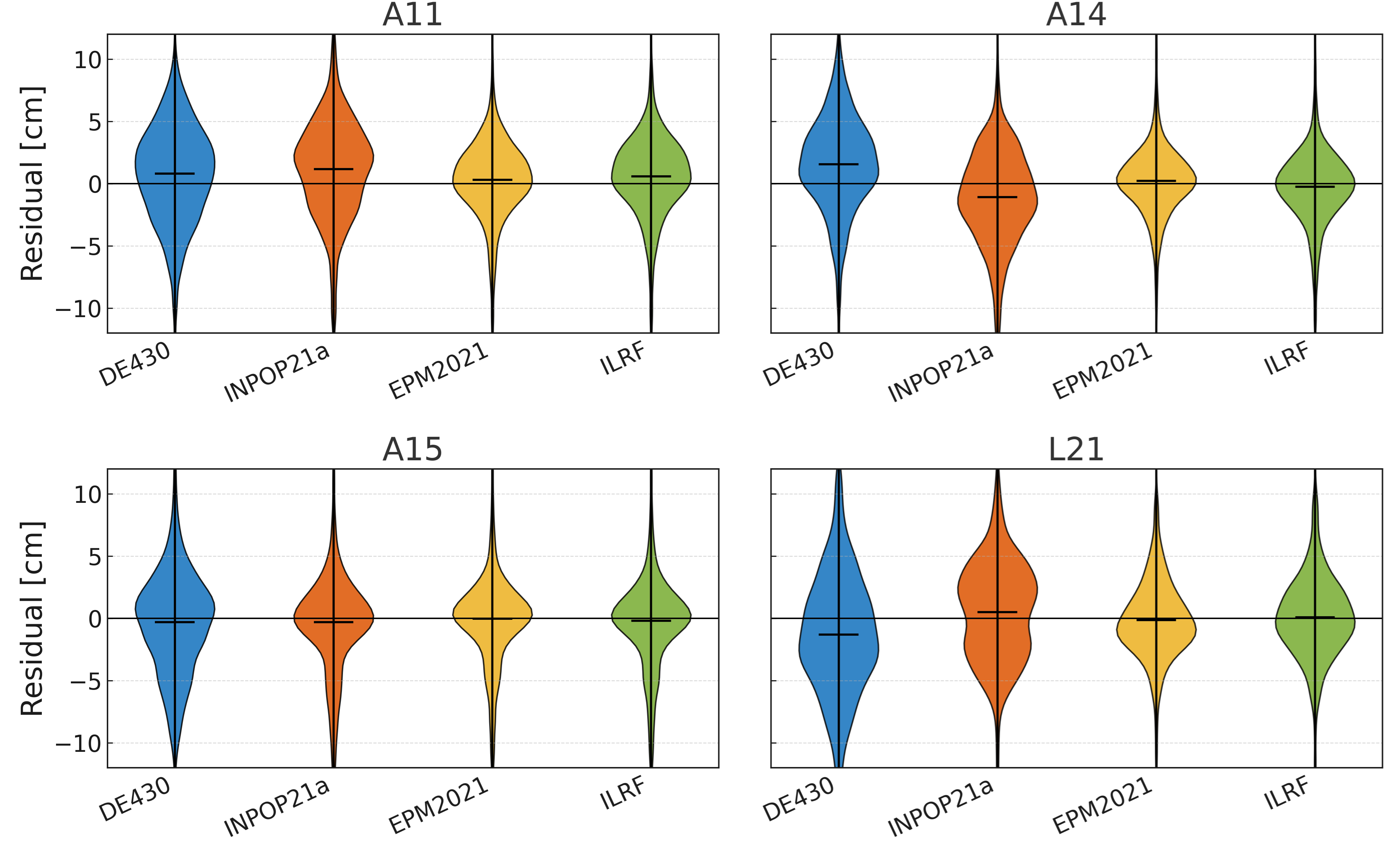}
      \caption{One-way LLR residuals derived using DE430, INPOP21a, EPM2021, and the combined ILRF without kinematic corrections for the period of 1984-2025.}
         \label{fig6old}
   \end{figure}

   \begin{figure*}[h!]
   \centering
   \includegraphics[width=1.0\hsize]{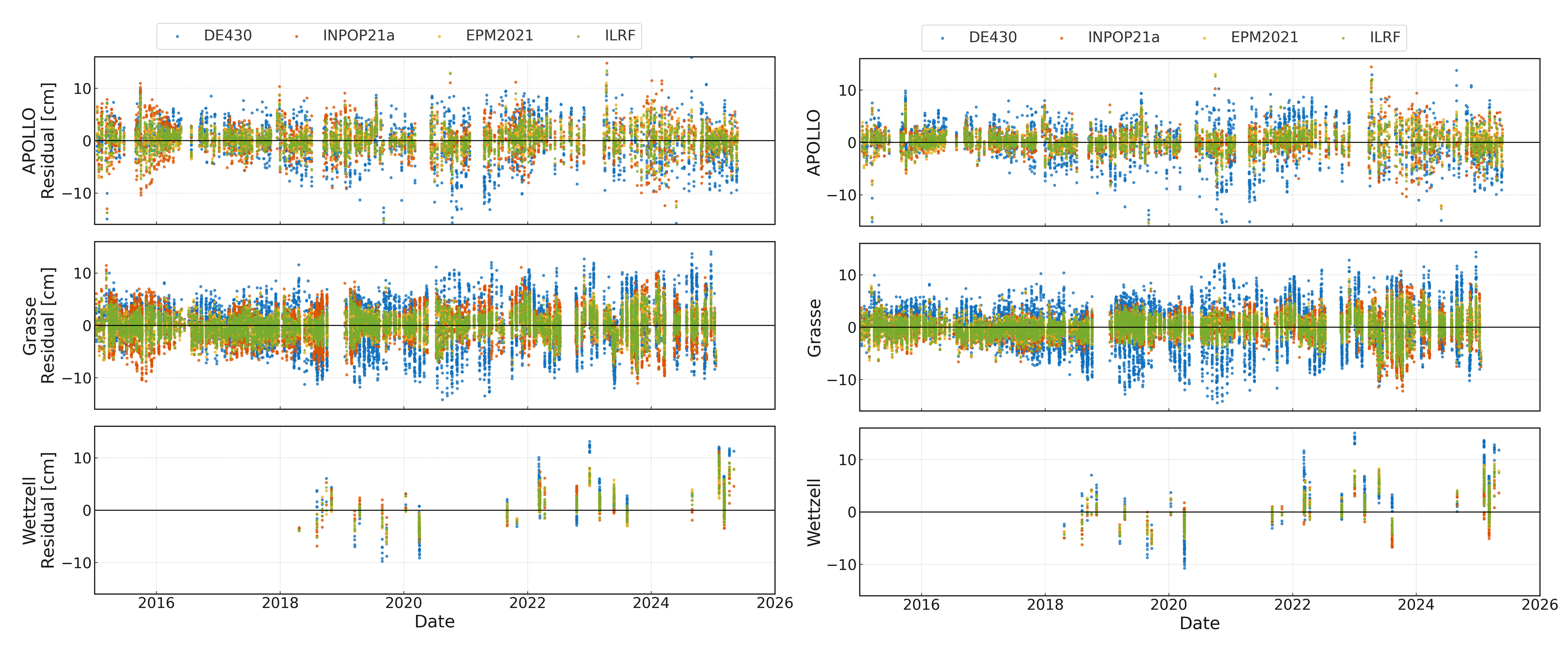}
      \caption{One-way LLR residuals derived using DE430, INPOP21a, EPM2021, and the combined ILRF without kinematic corrections (left) and with kinematic corrections (right) for the last 10 years with the highest quality of LLR observations. Please note that DE430 contains data until 2012, INPOP21a until 2020, and EPM2021 until 2021.}
         \label{fig6}
   \end{figure*}

\begin{table*}[ht]
\centering
\caption{RMS of LLR one-way residuals with kinematic corrections for selected high-performing stations in 1984-2025.}
\label{tab5a}
\begin{tabular}{lcccccc}
\hline\hline
RMS [cm]  & Grasse (g) & Grasse (IR) & APOLLO & Wettzell & Matera & McDonald \\
\hline
DE430     & 3.0 & 3.9 & 3.3 & 5.0 & 4.3 & 3.9 \\
INPOP21a  & 1.9 & 2.2 & 2.5 & 2.0 & 3.6 & 4.3\\
EPM2021   & 1.7 & 1.8 & 2.4 & 2.6 & 3.3 & 3.6\\
ILRF      & 1.7 & 1.8 & 2.3 & 2.5 & 3.3 & 3.6\\
\hline\hline
\end{tabular}
\end{table*}

LLR is the most accurate technique for lunar distance measurements, reference frame realizations, and thus also for the reference frame validation. First, we evaluate the ILRF using LLR without introducing any corrections or estimating any parameters, such as retroreflector positions, tidal displacements, or range biases. Figure \ref{fig6old}  shows the violin plots of one-way reflector-specific LLR residuals for DE430, INPOP21a, EPM2021, and the combined ILRF based on the whole time series of observations, whereas the raw station-specific LLR residuals are given in Fig. \ref{fig6} (left) for the last 10 years. Only the origin position and three Euler angles are considered, whereas the kinematic corrections are not applied. The ILRF is derived as a weighted mean of reference frames without the full integration for all solar system celestial bodies, as opposed to the contributing ephemerides. Despite that, the LLR residuals from the combined ILRF are close to the other solutions, oscillate around the zero with a spread not larger than for other ephemeris models. Figure \ref{fig6} proves the validity of the method for deriving the combined ILRF and the superior quality of the combination. Even without the full integration of motion for all celestial bodies, ILRF provides comparable quality to the best lunar ephemerides in terms of LLR residuals, especially in the most recent years.

\subsection{LLR validation with kinematic corrections}

\begin{figure}[h!]
   \centering
   \includegraphics[width=0.92\hsize]{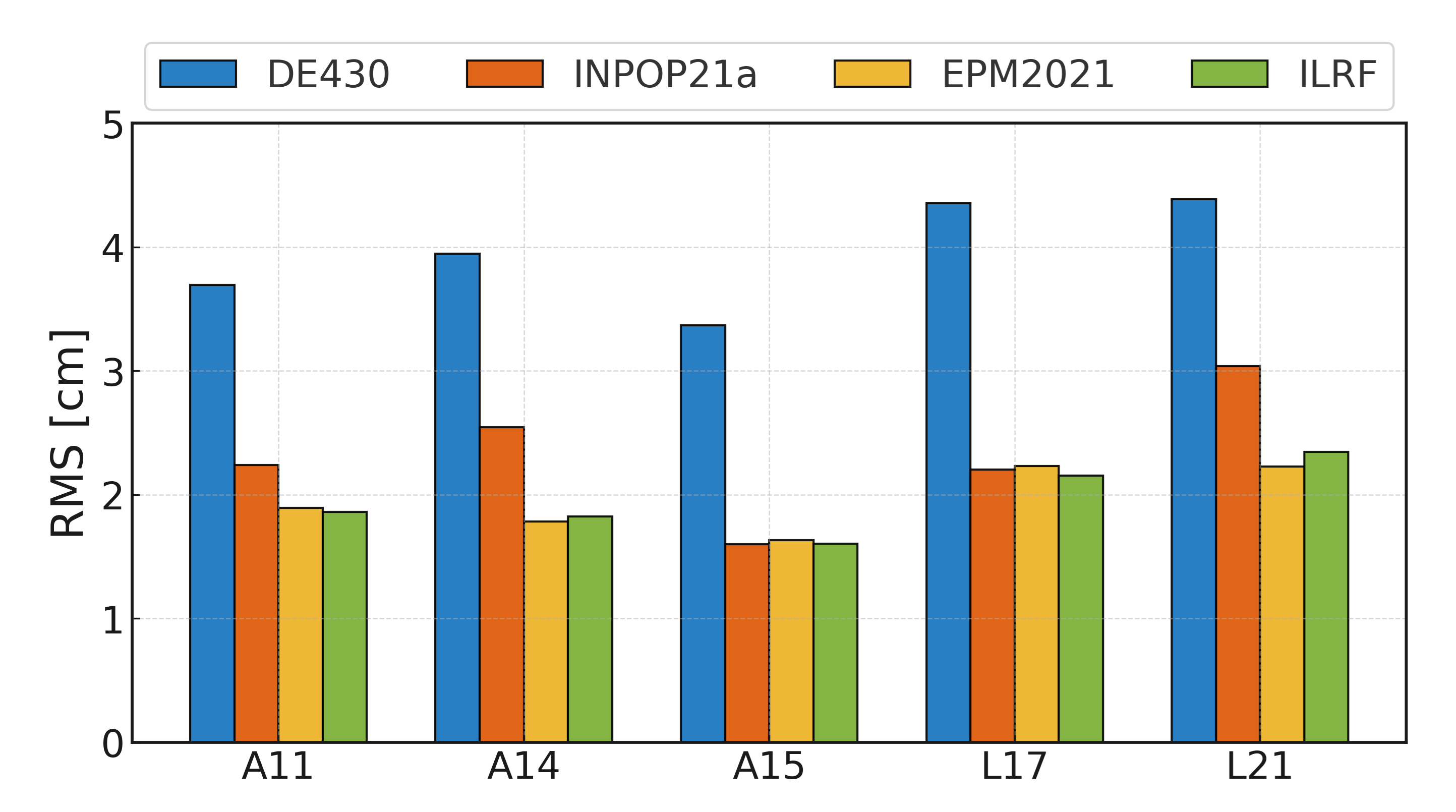}
      \caption{RMS of one-way LLR residuals for lunar retroreflectors when considering kinematic corrections for 2015-2025.  }
         \label{fig7}
   \end{figure}

Now, the LLR validation is conducted by considering the kinematic corrections for $A_1$, $A_2$, and $A_3$ parameters (see Section 4.2). The resulting RMS of residuals is equal to, e.g., 1.7, 1.8, 2.3, and 2.5~cm for Grasse (green), Grasse (IR laser), APOLLO, and Wettzell, respectively for the ILRF. The LLR residuals are at the level of 1.6-1.9~cm for Apollo retroreflectors and 2.2-2.3 cm for Luna 17 and 21 (see Fig. \ref{fig7}). These values allow ILRF to be classified as one of the best-performing models (see Table~\ref{tab5a} and Fig. \ref{fig6}, right). 

Please note that the RMS of LLR residuals at the level of 1.7-3.6 cm is much smaller than the mean overall errors of the ILRF at the level of 15.3 cm for the origin and 8.6 cm for the orientation. LLR residuals correspond, however, to the consistency of the radial component of the origin realization, which is equal to 1.2-2.1 cm for best-performing models after removing the mean offset (cf. Table \ref{tab2}). This again proves the high quality and sensitivity of LLR data to the radial component, but limited accuracy for constraining lateral components. The solution can be further improved by estimating range biases for LLR stations, which is avoided in this study because range bias estimates can absorb not only pure technical station circuit delays but also some geophysical processes.

\section{Transformation parameters}

\begin{figure}[h!]
   \centering
   \includegraphics[width=0.9\hsize]{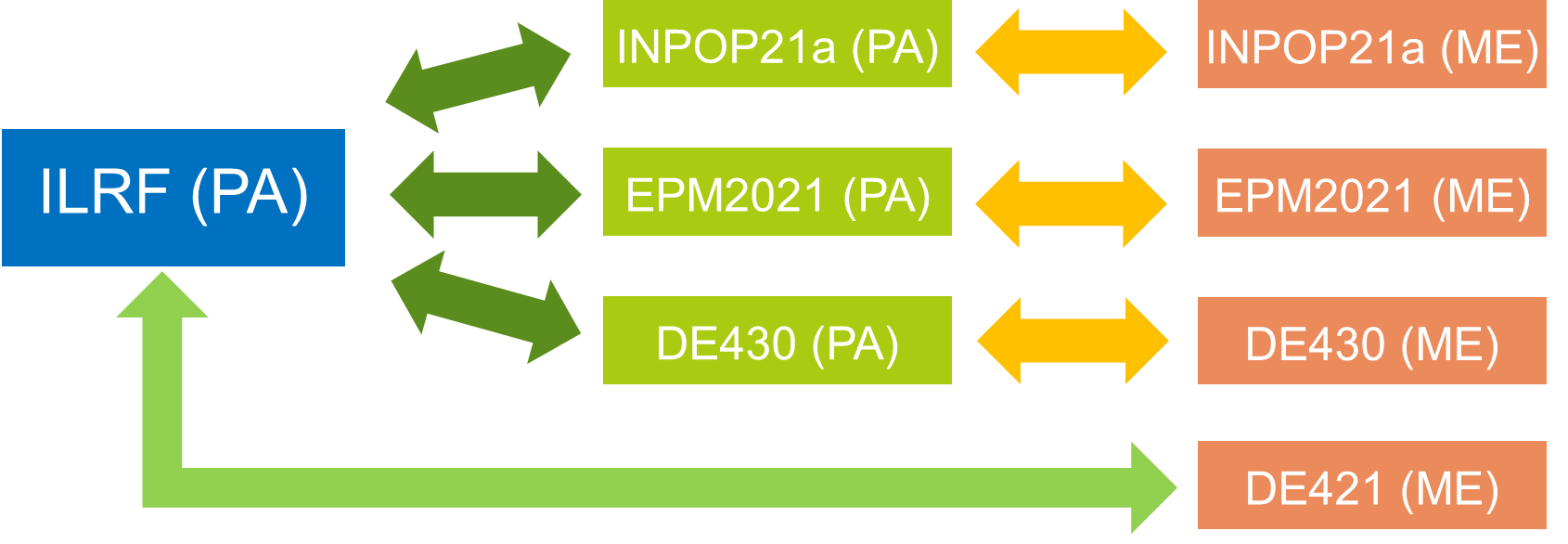}
      \caption{Transformation procedure between ILRF in PA, other PA lunar reference frames, and ME reference frames.}
         \label{fig8}
   \end{figure}

The transformation procedures and parameters are required to unify the lunar coordinates from different sources expressed in different reference frames and to keep the highest possible consistency between former, current, and future positions on the Moon or in the lunar orbit. Many cartographic products are provided in the ME frame, whereas positioning and navigation solutions are derived in PA. The proposed transformation procedure between ILRF and reference frames provided in ME includes two steps (for the full transformation) or one step (based on the Helmert transformation), as shown in Fig. \ref{fig8}.

\begin{table}[ht]
\centering
\caption{The positions of LLR retroreflectors in PA~ILRF. 
Please note that the positions of the Next Generation Lunar Retroreflector (NGLR-1) are preliminary 
due to the short time series of observations (since March 2025).}
\label{tab5}
\begin{tabular}{lrrr}
\hline\hline
Retroreflector & $X$ [m] & $Y$ [m] & $Z$ [m] \\
\hline
Apollo 11 & 1591966.745 &  690699.384 &   21003.764 \\
Apollo 14 & 1652689.627 & -520997.633 & -109730.514 \\
Apollo 15 & 1554678.397 &   98095.451 &  765005.257 \\
Luna 17   & 1114292.301 & -781298.502 & 1076058.718 \\
Luna 21   & 1339363.512 &  801871.855 &  756358.706 \\
NGLR-1    &  776672.915 & 1448359.255 &  552205.021 \\
\hline\hline
\end{tabular}
\end{table}

\begin{table*}[ht]
\centering
\caption{Transformation parameters between the combined ILRF and DE430, INPOP21a, and EPM2021 in PA, as well as DE421 ME derived using the retroreflector positions and their formal errors (1-$\sigma$). DE421 ME coordinates were taken from \citet{Williams2008}.}
\label{tab6}
\begin{tabular}{lrrrrrrr}
\hline\hline
ILRF vs. & $T_X$ [m] & $T_Y$ [m] & $T_Z$ [m] & $R_X$ [$10^{-6}$] & $R_Y$ [$10^{-6}$] & $R_Z$ [$10^{-6}$] & $Sc$ [$10^{-6}$] \\
\hline
DE430    & -0.1265 & -0.0580 &  0.1336 & -0.0089 & -0.0080 & -0.0630 &  0.0180 \\
$\pm$    &  0.0307 &  0.0199 &  0.0216 &  0.0211 &  0.0119 &  0.0053 &  0.0184 \\
\hline
INPOP21a & -0.0695 &  0.0248 & -0.0589 & -0.0010 &  0.0169 & -0.0321 & -0.0071 \\
$\pm$    &  0.0237 &  0.0154 &  0.0167 &  0.0163 &  0.0092 &  0.0041 &  0.0142 \\
\hline
EPM2021  &  0.1056 & -0.0001 & -0.0006 &  0.0039 & -0.0106 &  0.0501 &  0.0127 \\
$\pm$    &  0.0128 &  0.0083 &  0.0090 &  0.0088 &  0.0050 &  0.0022 &  0.0077 \\
\hline
ME DE421 & -0.1752 & -0.0144 &  0.1619 & -1.3539 & -381.3418 & -328.4958 &  0.1046 \\
$\pm$    &  0.0700 &  0.0456 &  0.0516 &  0.0432 &   0.0273 &   0.0206 &  0.0426 \\
\hline
ME DE421* & -- & -- & -- & -1.3596 & -381.2695 & -328.4838 & -- \\
$\pm$     & -- & -- & -- &  0.0625 &   0.0233 &   0.0222 & -- \\
\hline\hline
\end{tabular}
\end{table*}

All lunar reference frame realizations in PA are fixed to the lunar crust, use LLR data, and are based on the same concept of reference frame realization. Therefore, the expected differences between these realizations are rather small, as shown for the origin and orientation components. The LLR retroreflectors are the objects with positions derived with the highest accuracy, with the formal errors at the level of a few mm (1-$\sigma$) in particular solutions. Hence, we decided to employ the fitted retroreflector positions in each PA frame to derive the transformation parameters between ILRF and other ephemeris provided in PA. Table \ref{tab5} provides the VCE-based mean positions of retroreflectors in ILRF. The transformation formula based on the 7-parameter Helmert transformation reads as:

\begin{equation}
\left[\begin{matrix}X\\Y\\Z\\\end{matrix}\right]=\left[\begin{matrix}T_X\\T_Y\\T_Z\\\end{matrix}\right]+\left[\begin{matrix}1+Sc&R_Z&{-R}_Y\\-R_Z&1+Sc&R_X\\R_Y&-R_X&1+Sc\\\end{matrix}\right]\cdot\left[\begin{matrix}X_{ILRF}\\Y_{ILRF}\\Z_{ILRF}\\\end{matrix}\right]
\end{equation}
where $X, Y, Z$ on the left-hand side of the equation correspond to the resulting PA coordinates in DE430, INPOP21a, or EPM2021, whereas the translation $T_i$, rotation $R_i$, and the scale $Sc$ parameters are given in Table \ref{tab6}. The transformation parameters are estimated using the formal errors of the retroreflectors, which contribute to the variance-covariance matrix. Most of the rotation parameters, except for $R_Z$, are small and statistically insignificant (see Table~\ref{tab6}). The translation parameters are significant and may reach even 13 cm for $T_X$ and $T_Z$ components. Mean errors of deriving transformation parameters are equal to 3.5, 2.8, and 1.5~cm for DE430, INPOP21a, and EPM2021, respectively; therefore, the accuracy of the transformation procedure at the level of 2-3 cm is considered sufficient for most of the lunar applications. 

The inverse transformation procedure uses the same parameters and can be expressed as:

\begin{equation}
\left[\begin{matrix}X_{ILRF}\\Y_{ILRF}\\Z_{ILRF}\\\end{matrix}\right]=\left[\begin{matrix}1+Sc&R_Z&{-R}_Y\\-R_Z&1+Sc&R_X\\R_Y&-R_X&1+Sc\\\end{matrix}\right]^{-1}\left(\left[\begin{matrix}X\\Y\\Z\\\end{matrix}\right]-\left[\begin{matrix}T_X\\T_Y\\T_Z\\\end{matrix}\right]\right).
\end{equation}

\begin{figure}[h!]
   \centering
   \includegraphics[width=0.85\hsize]{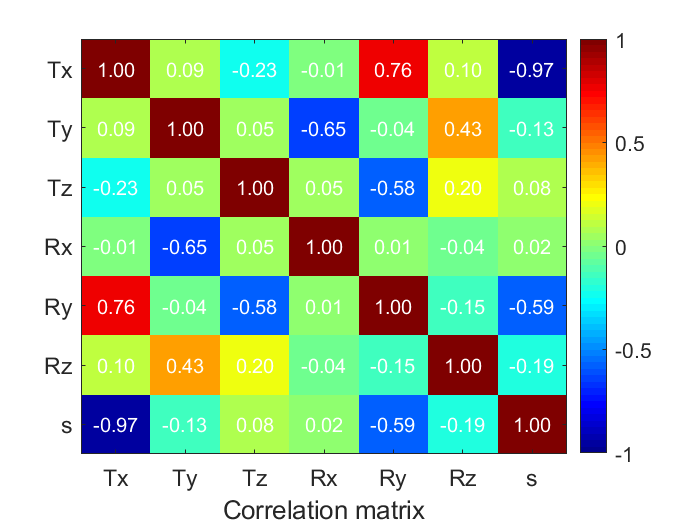}
      \caption{Correlation matrix of the transformation parameters between ILRF and other PA-based reference frames.}
         \label{fig9}
   \end{figure}

\begin{figure}[h!]
   \centering
   \includegraphics[width=1.0\hsize]{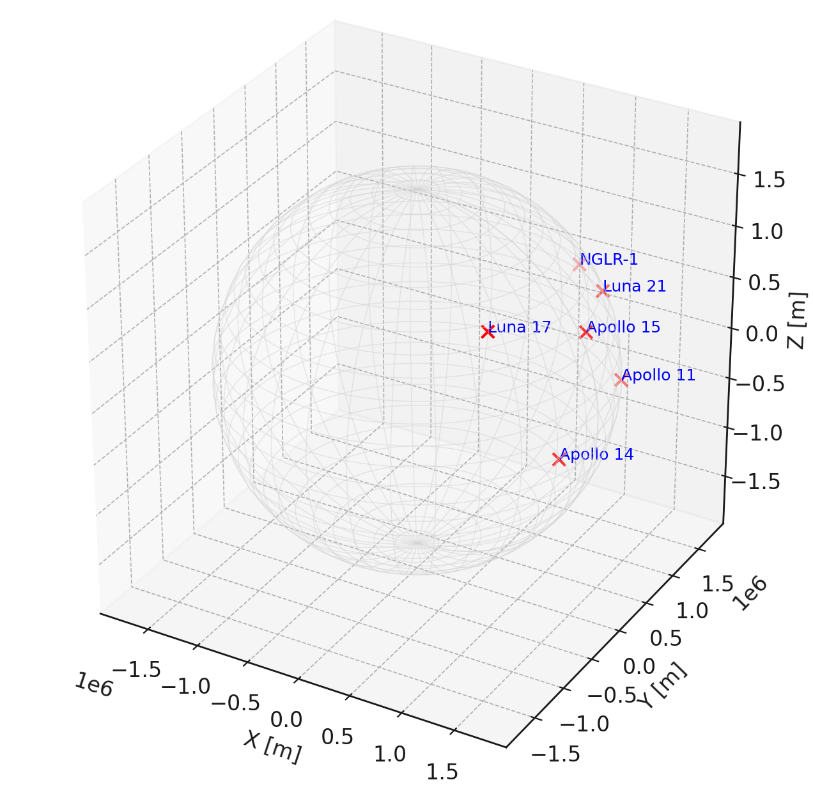}
      \caption{Distribution of the lunar retroreflectors on the lunar surface in ILRF (PA).}
         \label{fig10}
   \end{figure}

Interestingly, one of the largest transformation parameters between PA reference frames and the largest formal errors are obtained for the $T_X$ component, which is well observed by LLR because the position of the Earth coincides with the X-translation component. This can, however, be simply explained by the correlations between the $T_X$ component and the scale. Figure \ref{fig9} shows the correlation matrix between different translations, rotations, and the scale parameter. Due to the fact that the retroreflectors are placed only on the near side of the Moon and only one retroreflector is located in the southern hemisphere (see Fig. \ref{fig10}), the distribution of the retroreflectors does not properly sample the entire lunar surface, resulting in large correlations between transformation parameters. Despite that the X component is directly sensed by LLR, it becomes strongly correlated with the scale parameter and the correlation coefficient of $r=-0.97$. The discrepancies in Table \ref{tab6} for the $T_X$ component and the scale correspond well to the radial offset found in the combined origin of the ILRF. Thus, LLR can provide the positions of the retroreflectors on the lunar surface even with mm precision for the X component, but fails in reaching the actual center of mass of the Moon with an accuracy not better than 12 cm due to large correlations between the X component and the scale. This can, however, be improved by a better distribution of future retroreflectors or lunar orbiters with continuously tracked positions even above the far side of the Moon, which should enhance the observational techniques to reach the actual lunar center of mass.

The full transformation procedure between the PA and ME reference frames include five parameters and reads as:

\begin{equation}
ME=R_X\ \left(-p_{2c}\ \right)\ R_Y\left(p_{1c}\right){\ R}_Z\left(-\tau_c+\frac{I^2\sigma_c}{2}\right)PA,
\end{equation}
where $\tau_c$, $\sigma_c$ are libration angles, $I$ is the inclination of the equator of the Moon to the ecliptic of date, and ($p_{1c}$, $p_{2c}$) are components of the normal to the ecliptic in the PA reference system. 

A simplified transformation can be applied by neglecting some parameters or keeping only rotation angles as suggested by \cite{Park2021}. Table \ref{tab6} provides the transformation parameters between ILRF and DE421 ME in two versions – including the full set of 7 Helmert transformation parameters, and reducing the number of parameters to three rotation angles (denoted as DE421* ME in Table \ref{tab6}). DE421 ME has been selected because this reference frame is widely adopted for various products of lunar cartography.

The mean transformation error equals 5.2 and 9.5 cm for the 7-parameter transformation and the simplified 3-parameter transformation, respectively and differences between the two transformation procedures can reach up to 14 cm. Therefore, the 7-parameter transformation is recommended or the transformation based on eq. 17, to keep the accuracy at the highest level. Furthermore, considering precisely the definitions of the two moon-fixed frames, \cite{Rambaux2025} describe in detail transformation procedures for selected INPOP and DE ephemeris with a list of transformation parameters and their uncertainties, as well as the definition of the Lunar Time-scale and the time system transformation procedures. Thus, we refer to \cite{Rambaux2025} for details related to PA-ME transformations based on eq. 17.

\section{Discussion and conclusions}

We propose a definition for the Lunar Reference System and its first realization in the form of the International Lunar Reference Frame that includes the three ephemeris models for the Moon: DE430, INPOP21a, and EPM2021. Although the contributing models are based on LLR data, they are typically associated with different geophysical assumptions on the inner structure and elastic parameters or a different number of considered celestial bodies and values of fundamental constants, but also with different time coverage for the LLR data sample used for their fit. The resulting mean, combined model for the origin and orientation of the Moon is characterized by a similar accuracy to the best-performing lunar reference frame realizations. The weighting factors for ILRF are based on VCE, and the validity of the model covers the period 1970-2052. The combined ILRF averages intrinsically the background parameters, thus, benefits from the diversity of contributing models. 

The PA system has been selected as the reference system for ILRF orientation due to direct construction (or deduction) from lunar ephemerides, as well as the simplicity in terms of the transformation procedures. However, the transformation procedures between ILRF and various realizations derived in ME and PA are also possible via the provided transformation procedures. These procedures are useful for unifying the coordinates, maps, or cartography products historically provided in ME. The mean error of the transformation is 3.5 and 5.2 cm for the transformation from ILRF to DE430 PA and to DE421 ME, respectively, based on LLRs. The orientation parameters for the Moon expressed as three Euler angles are very well determined, resulting in a mean error of 8.2 cm for the period 1970-2052, which is remarkable considering the included 30-year prediction period.

The origin of the ILRF coincides with the center of mass of the Moon. However, the realization of the origin can be problematic because the LLR data are provided directly to the limited part of the lunar surface and not directly to the lunocenter. Hence, three issues of sensing the Moon’s center of mass were identified: (1) the radial component corresponding to the X-translation parameter is strongly correlated with the scale of the reference frame, (2) the along-track component loses its accuracy over time, especially after 2030, (3) the cross-track component shows periodical variations associated with the libration model errors. As a result, the realization of the ILRF origin is characterized by a mean error of 30.5 cm for the whole period and 15.3 cm for the period 2010-2030. The position of the lunocenter is thus materialized with the accuracy of one order of magnitude worse than the positions of the retroreflectors on the surface. The accuracy of the origin determination can be improved in the future by considering updated ephemeris based on a larger number of retroreflectors (especially in the southern hemisphere) or future lunar orbiters with high-quality orbits at the cm-level. Finally, the tie to the ITRF of LLR station is also of primary importance. Future observations of the LAGEOS-1/2 and LARES-2 satellites \citep{Sosnica2025} by LLR stations, and in particular by the APOLLO, will be crucial for improving the link between ITRF and ILRF. 

Other ILRF candidate solutions were tested, e.g., with a scale normalized to one of the contributing ephemeris, different sets of weights for the orientation and origin parameters, or using different ephemeris models. Separating the weights for the origin and orientation increased the LLR residuals by 2\%, whereas using DE440 instead of DE430 provided similar residuals for the origin component and increased the orientation residuals by a factor of five for precession and proper rotation parameters, possibly due to the extensive number of empirical kinematic parameters in DE440, which were not considered for the combination of the Euler angles. Therefore, DE430 was selected as the contributing model. The scale normalization leads to the same values of LLR residuals when estimating associated parameters and requires one additional transformation procedure; therefore, the scale normalization is not recommended to keep the highest possible simplicity of the solution. The geometrical scale issue disappears as soon as the positions of retroreflectors are provided in ILRF because the scale differences can entirely be absorbed by using the proper retroreflector coordinates.

The estimated total error of the ILRF realization is 17.6 cm for 2010-2030. However, the relative distances, selected components, or positions of the objects on the lunar surface can be derived with even greater accuracy. For example, the X component of the newly-installed retroreflector NGLR-1 estimated in ILRF is just a few mm due to the dominating LLR sensitivity to the X component of the lunar reference frame.

This paper describes partial outcomes from the IAG/IAU Joint Working Group 1.1.3 on Lunar Reference Frames. The future studies shall include the issues of lunar time systems, height systems, and reference points, as well as the Moon’s geoid model (or the selenoid model), all of which require unifications and standardizations to provide norms and standards for current and future lunar missions.

\begin{acknowledgements}
      The members of IAG and IAU JWG 1.1.3 on Lunar Reference Frames are acknowledged for providing valuable remarks to the combination results. We would like to thank the lunar ephemeris providers: DE, INPOP, and EPM, and the LLR stations for providing the very long series of high-quality lunar measurements. 
\end{acknowledgements}

%

\end{document}